\documentclass[sigconf, nonacm]{acmart}

\acmConference[RecSys '25]{19th ACM Conference on Recommender Systems (RecSys 2025)}{September 22–26, 2025}{Prague, Czech Republic}

\acmYear{2025}
\copyrightyear{2025}

\usepackage{subcaption} 
\usepackage{graphicx}

\AtBeginDocument{%
  }

\setcopyright{acmlicensed}
\copyrightyear{2025}
\acmYear{2025}
\acmDOI{XXXXXXX.XXXXXXX}
\acmISBN{978-1-4503-XXXX-X/2018/06}

\begin{document}

\title{Optimal signals assignment for eBay View Item page}

\author{Matan Mandelbrod}
\affiliation{%
  \institution{eBay BX-AI Research}
 \country{Israel}
}
\email{mmandelbrod@ebay.com}

\author{Biwei Jiang}
\affiliation{%
  \institution{eBay Buyer Success}
 \country{Israel}
}
\email{bjiang1@ebay.com}

\author{Giald Wagner}
\affiliation{%
  \institution{eBay BX-AI Research}
 \country{Israel}
}
\email{gwagner@ebay.com}

\author{Tal Franji}
\affiliation{%
  \institution{eBay BX-AI Research}
 \country{Israel}
}
\email{tfranji@ebay.com}

\author{Guy Feigenblat}
\affiliation{%
  \institution{eBay BX-AI Research}
 \country{Israel}
}
\email{gfeigenblat@ebay.com}

\begin{abstract}
Signals are short textual or visual snippets displayed on the eBay View-Item (VI) page, providing additional, contextual information for users about the viewed item. The aim in displaying the signals is to facilitate intelligent purchase and to incentivise engagement. 
In this paper, we present two approaches for developing statistical models that optimally populate the VI page with signals. Both approaches were A/B tested, and yielded significant increase in business metrics. 
\end{abstract}

\begin{teaserfigure}
  \centering
  \includegraphics[scale=0.3]{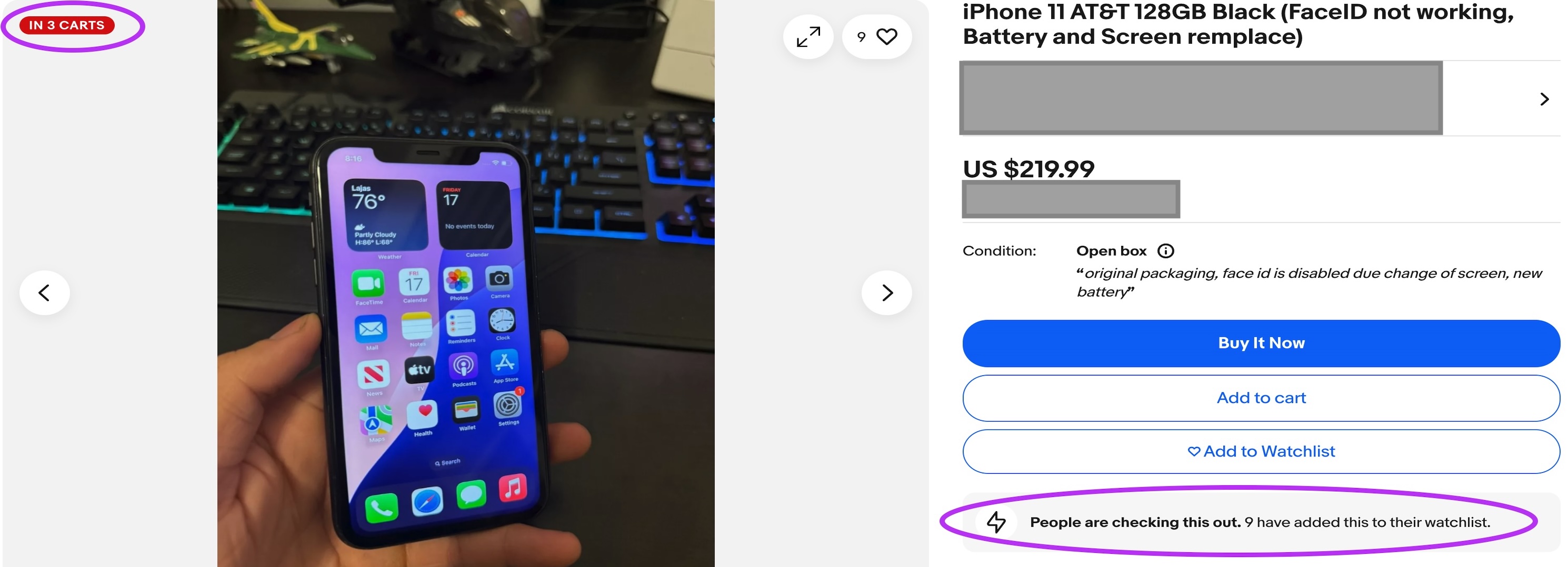}
  \caption{Urgency and conversational signals on eBay's VI page}
  \Description{Two signals displayed over difference placements marked in purple ovals}
  \label{fig:teaser}
\end{teaserfigure}


\maketitle


\section{Introduction} 
The View Item (VI) page is the most watched page in eBay, with hundreds of millions of daily impressions. This page displays all the information pertaining to the presented item, such as images, price, shipping, seller etc. In addition to this fixed data, optional textual and visual snippets, dubbed `signals' are displayed over predefined locations, or placements, on the VI page. Each placement has a fixed set of candidate signals that can potentially be displayed over it. From product perspective, signals address the difficulty buyers sometimes face in comparing items and finding relevant information about a specific item. Figure \ref{fig:teaser} shows signals displayed over two placements: Urgency over the picture panel, and Conversational below the engagement buttons (marked in purple ovals).

The aim of this work is to develop a statistical model which will determine how to optimally populate these placements with signals, where `optimally' is defined in terms of some business metric (e.g. purchase rate or GMB).  There are several placements on the VI page, each accommodates between one to two signals, selected from a designated set of signals for this placement. Thus a signal assignment for a VI impression consists of selecting a tuple of signals - one or two per placement. The work presented in this paper is limited to the Urgency and Conversational placements. 

Several challenges render this problem exceptionally demanding. Among them it is worth mentioning the small uplifts and the correlation between signals. The uplifts are the differences between the conversion rates when showing one signal instead of another. These are of order $10^{-4}$, hence the train and evaluation sets are required to be of order $O(10^8)$ so as to reduce overfitting and obtain statistically significant results. The correlation between signals refers to the fact that most of them are engagement indicators (number of purchases, number of addition to cart etc.), hence are typically varying together based on the items popularity.


The contributions described in this paper are as follows:
\begin{enumerate}
    \item We developed two ranking models for VI signals: Retrospective-learning (using XGBoost) and Conversion Likelihood Estimator. Both were A/B tested, and provided statistically significant performance improvements. Both models are fast and comply with the required SLAs (inference durations). The two approaches are discussed in section \ref{sec:approach}
   \item We developed an offline metric to evaluate the ranking models, which was essential to select the production models.
\end{enumerate}


\section{Related Work}

Existing literature on the subject is divided between behavioural and  algorithmic research. We'll refer here to the latter, beginning with \cite{cart_rf}, in which a reinforcement learning (RL) algorithm to reduce cart abandonment is designed and evaluated. The agent selects an action from a predefined set based on a features vector. Possible actions may include cart reorganization, discount offers and displaying scarcity messages. It was our decision, in order to facilitate fast go to market, to examine RL approaches such as contextual bandits (\cite{context_bandit}) in the next phases of development. Indeed, the signals assignment problem requires some adaptations related to the reward function (see discussion in \cite{rl_uplift} and \cite{uplift_bandit}), and handling of confounding variable, as well as development of online infrastructure for training and inference using RL.


In \cite{cart_forest} causal random forest is used to predict when (with versus without carts) and how  (scarcity versus price promotion) to target the pending shoppers. It uses demographic, historical purchase,  and product category features and achieves up to 29\% purchase rate increase when scarcity and urgency signals are displayed on the cart. Booking.com's paper \cite{goldenberg2020free} presents uplift modeling methods used in the context of coupon offering optimization. Though this problem differs from ours, the retrospective learning methods utilized in their paper resembles one of the approaches we adopted for our problem. We've experimented with several causal inference (\cite{pearl2016causal, hernan2020causal}) and uplift modeling (\cite{gutierrez2020causal}) as solution approaches (in particular, using  
 \cite{CausalML}). Since in our case, uplifts are very small, and since signals are highly correlated, this approach has not yielded (uplift modeling is typically used for drug treatment or for campaign targeting, such as coupons offering). We've therefore decided to concentrate on two different, simpler approaches as described below.

Finally, not directly discussing signals optimization, \cite{facebook_ads} compares observational studies with RCT (Randomized Controlled Trials) to measure Facebook's advertising effect. This paper provide good context for our decision to use an RCT in order to collect training data.

\section{Approach} \label{sec:approach}
We will now describe the two proven successful approaches. 

In the sequel we will be using the term `treatment' to indicate a particular signal assignment on the VI page. The outcome variable (related to the treatment), which will also be referred to as `conversion' is a binary engagement indicator, which obtains the value 1 for impressions that had engagement, and 0 otherwise. We used several different such engagement indicators: any BBOWAC \footnote{BBOWAC: Bin, Bid, Offer, Watch item, Ask the seller and Add to cart } action, bin or bid action, or purchase. Note that `conversion' is a generic term here, and does not necessarily refer to purchase. Finally, the notion `qualification set' refers to the subset of signals qualifying to be shown for a given VI impression. This is determined online based on predefined rules. 
\paragraph{Retrospective Learning } \label{retro_l}

The basic principle employed in this approach is to use actual conversions as features to train a classifier. In our settings, rather than using the signal as a feature in order to predict conversion, we used the conversion label as a training feature in order to predict the signal which is most likely to have been displayed given the conversion. Thus an input instance for the model training is a feature vector consisting of listing-related numerical features (such as item price, quantity sold, etc.), the binary qualification indicators (which signals qualified for this impression), and the conversion indicator. The training labels are the ids of the shown signals for this impression, so that the model learns to predict the signal most likely to have yielded the conversion (either 0 or 1). Upon inference the value of the conversion feature is set to 1 and concatenated to the contextual features. The idea is for the model to predict the signal assuming conversion will take place. 



In contrast to \cite{goldenberg2020free}, which uses only the positive examples to learn the conversion probability, our experiments suggest that adding an equal number of negative samples to the training data significantly improved the offline metrics. One possible explanation would be that since the uplifts are very small, incorporating both negative and positive examples has some amplification effect of the slight uplift patterns. The merit of this approach is the fact that it reduces the hard uplift problem (hard due to the tiny uplifts) to a standard classification problem, in which the weak treatment feature turns to be the label for prediction. 

\paragraph{Conversion Likelihood Estimator} \label{direct_stats}
In this approach, we directly calculate the signals conversion rates over each qualification set, and rank them (within the qualification set) based on these rates, provided their difference (which is the uplift) is statistically significant. This requires a large amount of data per qualification set, in order to obtain a statistically significant result. This simple approach can be thought of as a classifier with only the qualification indicators as features. That is, per each qualification tuple the model assigns the ranking for this tuple. 

\subsection{Offline Metrics}
Offline metrics are required in order to measure the performance of various models prior to testing online. In our case, the choice of an appropriate metric is not obvious. Seemingly, at least for \nameref{retro_l}, metrics should be trivial, since the model is basically a classifier, so AUC, F1 or any other standard metric should work. But it turns out that this is not the case: first, as described above, the uplifts are extremely small, and the different signals are barely separable, resulting in AUCs of $O(10^{-3})$ above random (recall that this does translate to a significant GMV uplift).
Second - the classification metrics may be fair (on the scale aforementioned), but yield negative uplifts, since the model mistakenly predicts a signal of lower conversion rate, due to poor initial splitting or overfitting (which can frequently occur due to the slight differences in conversion rates), and the standard classification metrics fail to indicate that the predicted signals do not increase conversion rates.
Uplift metrics seem naturally suitable for the problem at hand. Several uplift metrics are used in the literature (\cite{gutierrez2020causal}), and they enable comparison between different uplifts models. However, they are mainly designed  towards quantifying the treatment cost v.s. uplift tradeoff in order to identify e.g. the campaign budget cutoff point. In our case, this is not relevant, as showing signals doesn't require any budget, and we are purely interested in the expected outcome (i.e. conversion rates) of applying one ranking model v.s. another. 

\begin{figure}[h]
    \centering
    \subfloat[Randomly shown in RCT\label{fig:rand_201_202}]{%
        \includegraphics[width=0.25\textwidth]{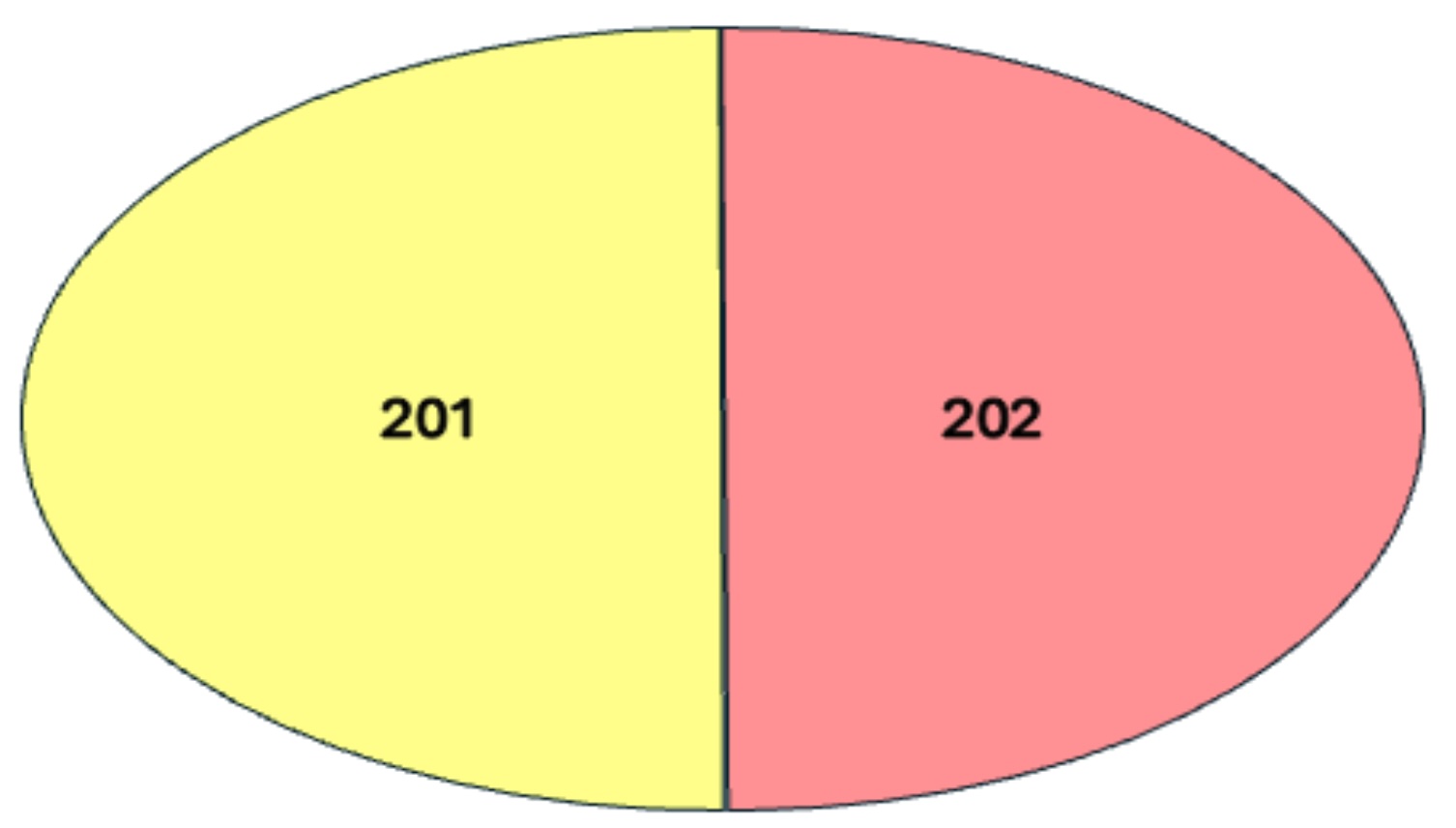}
    }   
    \subfloat[Model's prediction overlaid over actual\label{fig:model_201_202}]{%
        \includegraphics[width=0.25\textwidth]{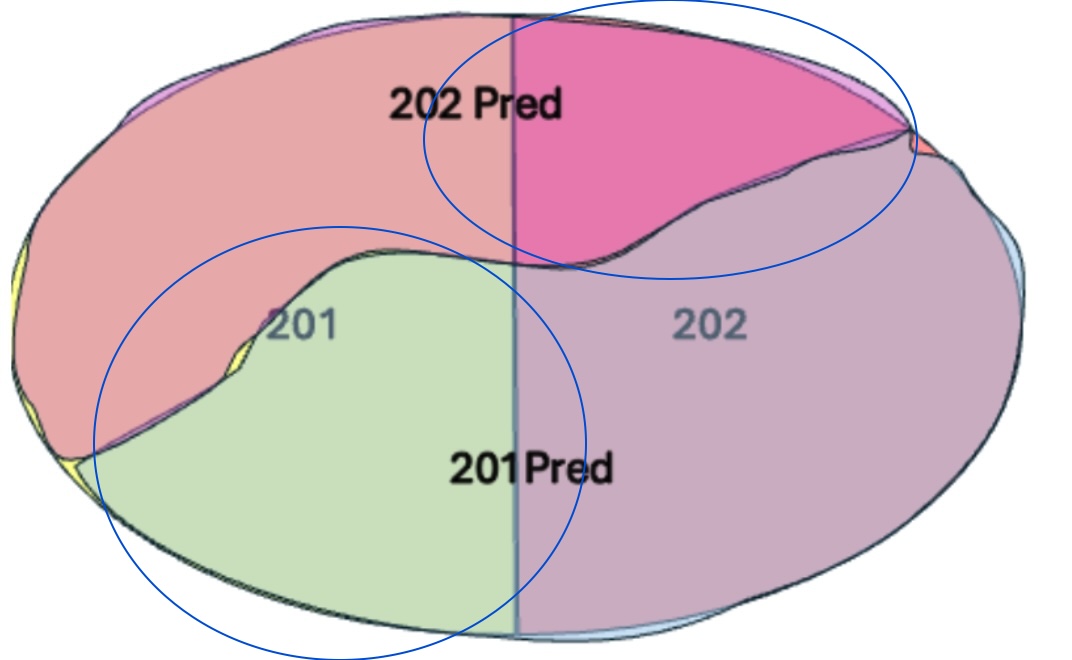}
    } \\
    
    \subfloat[Full randomization\label{fig:general_rand}]{%
        \includegraphics[width=0.25\textwidth]{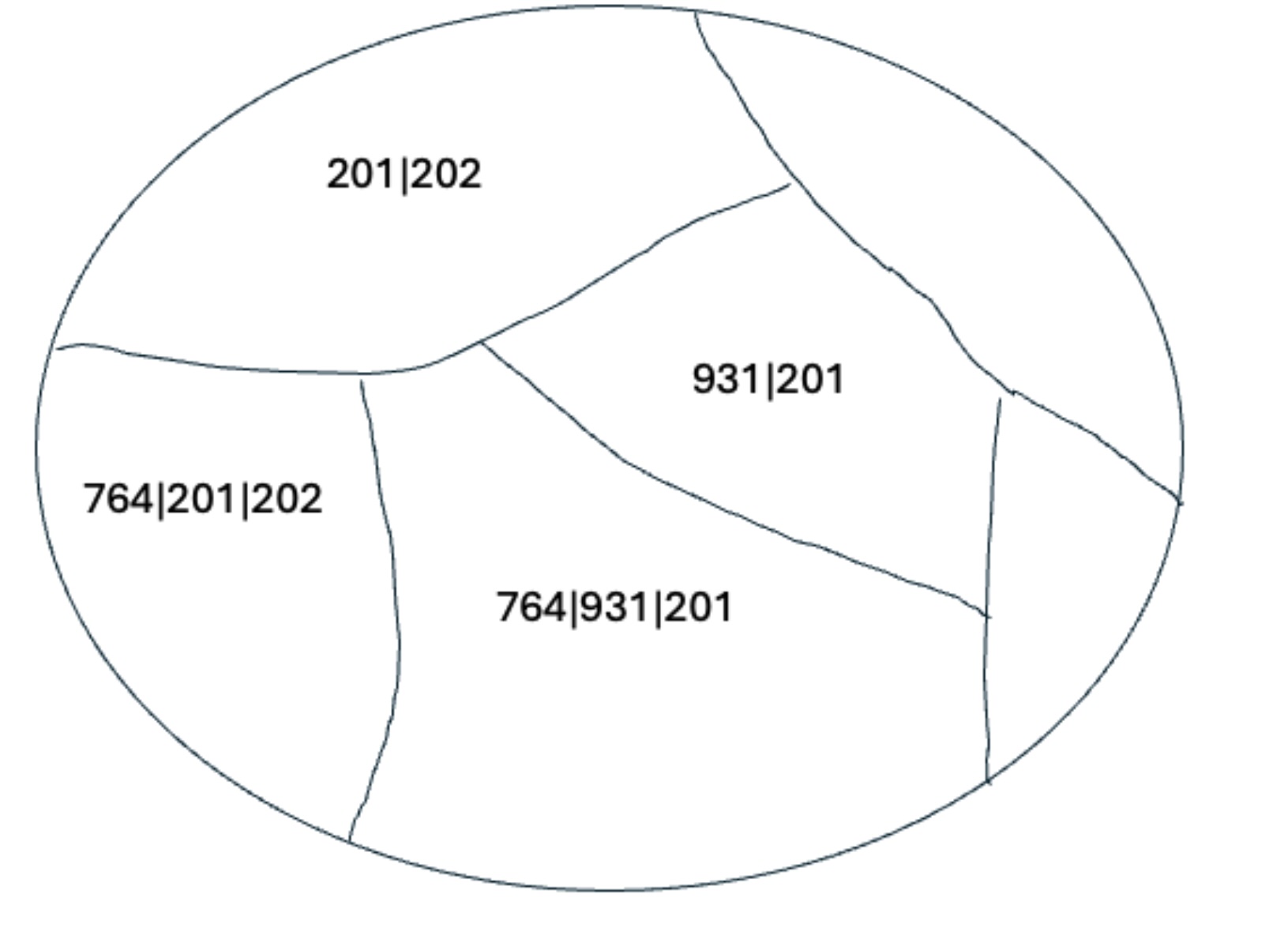}
    }
    \subfloat[Overlaid with model-induced prediction\label{fig:general_rand_and_prod}]{%
        \includegraphics[width=0.23\textwidth]{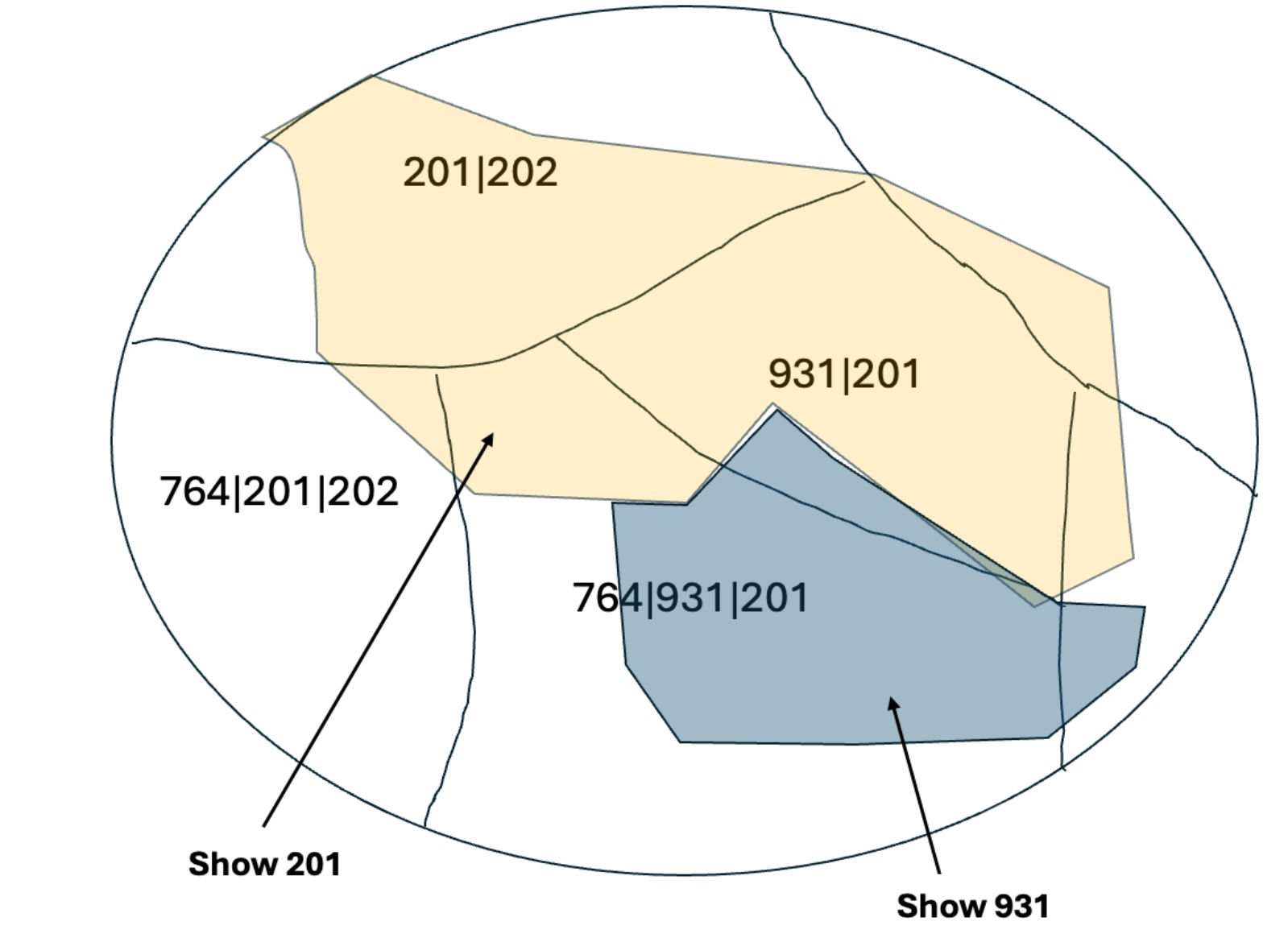}
    }
    
    \caption{Illustration of shown signals in randomization and prediction}
     \label{fig:full_figure_2} 
\end{figure}

To this end, we developed a designated offline metric that directly estimates the uplift of the model's assignment. The estimator is based on the adjustment formula \cite{pearl2016causal}. The training is done using data collected from a \textit{conditional randomization} (see \cite{hernan2020causal}, pp. 17), meaning that within each qualification set signals are assigned randomly.
Let us assume a set of only two qualified signals: $s_1, s_2$ randomly shown to users, and denote by $V_1, V_2$ the subsets consisting of units (VI impressions) exposed to $s_1, s_2$ respectively, that is - impressions in which users viewed these signals.  The sizes of these sets are equal since the two signals are randomized. We further assume both signals qualified for all impressions. This situation is illustrated in fig. \ref{fig:rand_201_202}: half the population viewed signal 201 and the other half viewed 202. Let us further assume a signals ranking model $M$ over this set. By applying the model on all units we obtain an induced partition of the entire units set into $V^M_1, V^M_2$, where $V^M_i$ is the subset of units for which the model $M$ predicted "show signals $s_i$". We can empirically estimate the actual conversion probabilities (or rates) of each signal using the randomized data as $$P(Y = 1 | T = s_i) \approx \frac{\sum_{V_i} Y}{N_i}, $$ where $Y$ is the conversion indicator (e.g. 1 if any BBOWAC took place, 0 otherwise), $T$ is the treatment variable (which, in our example, can obtain one of the values $s_1, s_2$), and $N_i$ is the size of $V_i$. This is simply the empirical mean of the conversion indicators over each subset. But what would have been the conversion probabilities had signals assignment followed the model's prediction rather than the random one? Since in reality this assignment never took place, the estimand in this case is referred to as "the counterfactual conversion probability". We will estimate it by answering the question: "what would have been the conversion rate over $V^M_i$ had all units in this subset were shown $s_i$?". The mathematical notation for this quantity is expressed using the \textbf{do} operator (\cite{pearl2016causal}), and estimated, in this simple case as: 
\begin{equation}
    \label{basic_do}
    P(Y=1 | do(T=s_i), V^M_i) = P(Y=1 | T=s_i, V^M_i) \approx \frac{\sum_{V_i \cap V^M_i}Y}{| V_i \cap V^M_i |}.
\end{equation}  In words: the estimation of the conversion rate if all users in $V_i^M$ had been exposed to $s_i$ is just the conversion probability over the set that both actually saw $s_i$ and has been assigned $s_i$ by the model. Here $|A|$ is the size of the set $A$, and the reason that $do(T=s_i)$ translates to $T=s_i$ in eq. \ref{basic_do} is that there are no confounding variables in this simple scenario. This is illustrated in fig. \ref{fig:model_201_202} - the shaded areas are the partition induced by the model, and the areas marked by the blue ovals are the ones used to calculate the estimated conversion rates of the predicted assignment. Averaging over both assignment sets provides an estimate for the total counterfactual conversion rate:
$$ \hat{c} \approx \sum_i \frac{|V_i^M|}{N} \frac{\sum_{V_i \cap V^M_i}Y}{ | V_i \cap S^V_i |}. $$

In the general case, when we calculate over the entire tests set, we need to adjust for the confounding variables (the qualification indicators): 

\begin{align}
\label{do_z}
P(Y=1 \mid do(T=s_i), V^M_i) 
&= \sum_Z P(Y=1 \mid T=s_i, Z=z, V^M_i) \notag \\
&\quad \cdot P(Z=z \mid V_i^M)
\end{align}

where $Z$ is the qualification indicator variable. The estimate of the overall counterfactual conversion rate is obtained by averaging eq. \ref{do_z} over all prediction sets $V^M_i$: 
$$
 \hat{c} \approx \sum_i \frac{|V_i^M|}{N} P(Y=1 | do(T=s_i), V^M_i).
$$
This is illustrated in Figure ~\ref{fig:full_figure_2}. The partition in Fig. \ref{fig:general_rand} shows the different qualification sets of the randomized data. Within each such set, signals are randomly shown to users in the VI impression. The partition induced by the prediction of the models is overlaid in ~\ref{fig:general_rand_and_prod} as the colored patches that span several qualification sets. The adjustment formula eq. \ref{do_z} is applied within each such patch to estimate the counterfactual conversion rate, and the overall estimate is obtained by averaging across all the colored patches. 
 It is worth noting that we have not encountered such a metric in the literature, which directly estimates the effect of applying a model over an existing dataset.

\section{Results}

The Retrospective Learning model yielded consistent uplifts across several buyer engagement metrics, such as Add to Cart (+0.36\%) and Best Offer (+0.49\%). The Conversion Likelihood Estimator showed a statistically significant uplift in GMB. It also contributed increase in Organic Revenue. Overall the signal ranking models not only drive GMB, but also improve user engagement throughout the entire purchase journey—from top-funnel impression uplift on the View Item page, to mid-funnel BBOWAC actions, and ultimately to bottom-funnel. We conjecture that GMB uplifts could be achieved by the retrospective learning model by increasing the weights of conversion related indicators at training, and possibly by adjusting the loss function (this is a WIP)






\bibliographystyle{ACM-Reference-Format}
\bibliography{sample-base}



\end{document}